\documentclass[a4paper,reprint,12pt,aps,tightenlines,onecolumn,notitlepage,superscriptaddress]{revtex4-1}
\usepackage{amsmath}
\usepackage{amssymb}
\usepackage{cancel}
\usepackage{stmaryrd}
\usepackage{amsfonts}
\usepackage[T1]{fontenc}
\usepackage{bm}
\usepackage{color}
\usepackage{graphicx}
\DeclareSymbolFont{operators}{OT1}{cmr}{m}{n}
\DeclareSymbolFont{letters}{OML}{cmm}{m}{it}
\DeclareSymbolFont{symbols}{OMS}{cmsy}{m}{n}
\DeclareSymbolFont{largesymbols}{OMX}{cmex}{m}{n}
\usepackage[hyperindex,breaklinks]{hyperref}
\hypersetup{
    pdfborder={0 0 0},
    allbordercolors={0 0 0},
    bookmarks=true,         
    unicode=false,          
    pdftoolbar=true,        
    pdfmenubar=true,        
    pdffitwindow=false,     
    pdfstartview={FitH},    
    pdfcreator={Eero Hirvijoki},
    pdfproducer={pdflatex},
    pdfnewwindow=true,
    colorlinks=true,
    linkcolor=blue,
    citecolor=blue,
    filecolor=blue,
    urlcolor=blue,  
}
\usepackage{array}

\newcommand{\fa}{\mathcal{A}}
\newcommand{\fb}{\mathcal{B}}

\newcommand{\fe}{\mathcal{E}}

\newcommand{\fn}{\mathcal{N}}
\newcommand{\fs}{\mathcal{S}}
\newcommand{\fp}{\mathcal{P}}
\newcommand{\fu}{\mathcal{U}}

\makeatother
\begin{document}

\title{Energy-, momentum-, density-, and positivity-preserving spatio-temporal discretizations for the nonlinear Landau collision operator with exact H-theorems}

\author{Eero Hirvijoki}
\affiliation{Princeton Plasma Physics Laboratory, Princeton, New Jersey 08543, USA}
\email{ehirvijo@pppl.gov; eero.hirvijoki@gmail.com}

\author{Joshua W. Burby}
\affiliation{Courant Institute of Mathematical Sciences, New York, New York 10012, USA}
\email{joshua.burby@cims.nyu.edu}

\author{Michael Kraus}
\affiliation{Max-Planck-Institut f\"ur Plasmaphysik, Boltzmannstra\ss e 2, 85748 Garching, Deutschland}
\email{michael.kraus@ipp.mpg.de}
\affiliation{Technische Universit\"at M\"unchen, Zentrum f\"ur Mathematik, Boltzmannstra\ss e 3, 85748 Garching, Deutschland}
\email{michael.kraus@ipp.mpg.de}

\date{\today}

\begin{abstract}
This paper explores energy-, momentum-, density-, and positivity-preserving spatio-temporal discretizations for the nonlinear Landau collision operator. We discuss two approaches, namely direct Galerkin formulations and discretizations of the underlying infinite-dimensional metriplectic structure of the collision integral. The spatial discretizations are chosen to reproduce the time-continuous conservation laws that correspond to Casimir invariants and to guarantee the positivity of the distribution function. Both the direct and the metriplectic discretization are demonstrated to have exact H-theorems and unique, physically exact equilibrium states. Most importantly, the two approaches are shown to coincide, given the chosen Galerkin method. A temporal discretization, preserving all of the mentioned properties, is achieved with so-called discrete gradients. Hence the proposed algorithm successfully translates all properties of the infinite-dimensional time-continuous Landau collision operator to time- and space-discrete sparse-matrix equations suitable for numerical simulation.
\end{abstract}

\maketitle

\section{Introduction}
\label{sec:introduction}
This paper is devoted to investigating the discretization of the collisional relaxation problem encountered in plasmas. There, a distribution function $f(\bm{v},t):\mathbb{R}^3\times\mathbb{R}_{\geq 0}\mapsto \mathbb{R}_{\geq 0}$ is assumed to evolve according to the equation
\begin{align}\label{eq:landau}
\frac{\partial f}{\partial t}=\frac{\partial}{\partial \bm{v}}\cdot\int \limits_{\mathbb{R}^3}\mathbb{Q}(\bm{v}-\bm{v}')\cdot\left( f(\bm{v}')\frac{\partial f}{\partial\bm{v}}-f(\bm{v})\frac{\partial f}{\partial \bm{v}'}\right)d\bm{v}',
\end{align}
corresponding to the dynamics driven by the nonlinear Landau collision operator~\cite{Landau:1936}. The dyad $\mathbb{Q}(\bm{\xi})=(\mathbb{I}-\bm{\hat{\xi}}\bm{\hat{\xi}})/|\bm{\xi}|$ in the above expression is an inversely scaled projection matrix with an eigenvector $\bm{\xi}$ corresponding to zero eigenvalue, and $\bm{\hat{\xi}}=\bm{\xi}/|\bm{\xi}|$. For practical purposes, the discussion in this paper will be limited to the single-species collisional relaxation problem, and we will consider only the velocity-space evolution, using normalized units to avoid unnecessary clutter. Nothing, however, prevents generalizing our results to multiple species, if need be.

The above collisional relaxation problem can be formulated in a weak sense, given an arbitrary time-independent test function $u(\bm{v})$. The weak formulation is
\begin{align}\label{eq:weakform}
\frac{d}{dt}M(u, f)=C_f(u , \ln f),
\end{align}
where the symmetric, bilinear forms $M$ and $C_f$ are defined according to
\begin{align}
M(g,h)&=\int \limits_{\mathbb{R}^3} g h\,d\bm{v},\\
C_f(g,h)&=-\frac{1}{2}\iint \limits_{\mathbb{R}^3\times\mathbb{R}^3}\left(\frac{\partial g}{\partial \bm{v}}-\frac{\partial g}{\partial \bm{v}'}\right)\cdot f(\bm{v})\mathbb{Q}(\bm{v}-\bm{v}')f(\bm{v}')\cdot\left(\frac{\partial h}{\partial \bm{v}}-\frac{\partial h}{\partial \bm{v}'}\right) d\bm{v}d\bm{v}'.
\end{align}
Provided that the distribution function $f$ is nonnegative, the form $C_f(u,w)$ is negative semidefinite, with a left-right null-space $\phi=\{|\bm{v}|^2,\bm{v},1\}$.

Using the weak formulation, it is straightforward to show that the functions $\phi(\bm{v})=\{|\bm{v}|^2,\bm{v},1\}$ generate invariant forms $M(\phi,f)$ with respect to the dynamics and correspond to the conservation laws of energy, momentum and density. These follow directly from the null-space of $C_f(u,w)$. Consequently, the condition $C_f(u,\ln f)=0$, with respect to arbitrary $u$, requires that $\ln f$ is a linear combination of the functions $\phi=\{|\bm{v}|^2,\bm{v},1\}$, corresponding to a Maxwellian equilibrium state. Furthermore, because $M(1,f)$ is an invariant and $C_f(u,u)\leq 0$, one finds that $\partial_t M(-\ln f,f)\ge 0$, which completes the H-theorem (monotonic entropy production that vanishes only for the equilibrium state). The final property of the Landau collision integral is that it preserves the positivity of the distribution function: assuming $f$ to be at least twice differentiable and non-negative, then, at a point $\bm{v}^{\star}$ where $f(\bm{v}^{\star})=0$, $\partial_{\bm{v}}f(\bm{v}^{\star})=0$, and $\partial^2_{\bm{v}\bm{v}}f(\bm{v}^{\star})$ is positive semi-definite, the evolution equation provides
\begin{align}
\frac{\partial f(\bm{v}^{\star})}{\partial t}=\int\limits_{\mathbb{R}^3}\mathbb{Q}(\bm{v}^{\star}-\bm{v}')f(\bm{v}')d\bm{v}':\frac{\partial^2f(\bm{v}^{\star})}{\partial\bm{v}\partial\bm{v}}\geq 0.
\end{align}

With this paper, we target spatio-temporal discretization methods that preserve the above mentioned properties exactly, to machine precision. We shall take two different routes which, in the end, are shown to coincide. In Sec.~\ref{sec:direct-Galerkin}, we propose a Galerkin discretization of the weak form~\eqref{eq:weakform} that ``accidentally'' succeeds in achieving all the desired properties in a spatially discrete but time-continuous system. In Sec.~\ref{sec:metriplectic-structure}, we provide a mathematical explanation for this accident, demonstrating that the proposed spatial Galerkin discretization in fact defines a finite-dimensional metriplectic structure (for an introduction to metriplectic dynamics, see~\cite{Morrison:1986vw,Grmela:1984dn}). In Sec.~\ref{sec:metriplectic-discretization}, we demonstrate how the very same spatial discretization is obtained directly from the underlying infinite-dimensional metriplectic structure of the Landau collision operator. Finally, in Sec.~\ref{sec:temporal}, we propose a temporal integration method based on the concept of discrete gradients~\cite{Quispel:1996,McLachlan:1999,CohenHairer:2011} which is shown to preserve all the desired properties to machine precision. The paper is concluded in Sec.~\ref{sec:discussion}.

\section{Direct Galerkin discretization}
\label{sec:direct-Galerkin}
In the past and present, direct Galerkin methods for the Landau collision operator have received and are receiving significant attention from applied mathematicians~\cite{Buet_LeThanh:hal-00092543,Yoon:2014:POP,Taitano:2015JCP,Hager:2016:JCP}. Based on our previous work~\cite{Hirvijoki:2017ei}, we have learned that straightforward Galerkin discretizations of the form 
\begin{align}\label{eq:old-discretization}
f_h(\bm{v},t)=\sum_if^i(t)\psi_i(\bm{v}),
\end{align}
manage to preserve the conservation laws exactly, as long as the basis $\{\psi_i\}_{i\in I}$ is capable of representing quadratic functions exactly within the domain of support for the chosen basis. The conservation laws are thus somewhat trivial to achieve with polynomial second-order finite-element methods. Unfortunately, the preservation of positivity and the existence of an H-theorem are much trickier. Discretizations of type~\eqref{eq:old-discretization} in general cannot guarantee the strict non-negativeness of $f_h$. This quickly turns into realizability issues in simulations and is against the basic principles of physics. %The existing positivity-preserving methods, on the other hand, are based on brute-force, so-called flux-limiting finite-volume discretizations~\cite{Vanleer:1977,Gaskell_Lau:1988}, and are typically incapable of reproducing the conservation laws without additional \textit{ad hoc} modifications of the original equation~\cite{Taitano:2015JCP}. Finally, none of the existing methods provide an H-theorem for the temporally discretization.
After some reflection, we have found a simple, yet elegant solution to the positivity-preservation problem that happens to be consistent with conservation laws and the H-theorem. The recipe is simple -- the root idea can be tracked down to the logarithm present in~\eqref{eq:weakform}. This section is devoted to describing the recipe in detail.

We begin by choosing an \textit{ab initio} positive discretization
\begin{align}\label{eq:new-discretization}
f_h=\exp(g_h), && g_h=\sum_{i\in I} g^i(t)\psi_i(\bm{v}),
\end{align}
with $\{\psi_i\}_{i\in I}$ a second order Galerkin basis with compact support, and $\{g^i\}_{i\in I}$ the degrees of freedom for $g_h$. Choosing a test function $u=\psi_i$, direct substitution of~\eqref{eq:new-discretization} to the weak formulation~\eqref{eq:weakform} provides us with a nonlinear matrix equation
\begin{align}\label{eq:Galerkin-matrix}
\sum_{j\in I} M(\psi_i, f_h \psi_j) \frac{dg^j}{dt} = \sum_{j\in I} C_{f_h}(\psi_i , \psi_j)g^j \quad \forall\, i\in I,
\end{align}
which is a linearly-implicit expression for the equations of motion of the degrees of freedom. Here, the integrals within the forms $M$ and $C_f$ are naturally limited to the domain of support for the basis. Also, note that the square matrices $M(\psi_i, f_h \psi_j)$ and $C_{f_h}(\psi_i , \psi_j)$ depend on the degrees of freedom via $f_h$, and that while $M(\psi_i, f_h \psi_j)$ is sparse, $C_{f_h}(\psi_i , \psi_j)$ is not. Equation~\eqref{eq:Galerkin-matrix} will be our work horse throughout the rest of the paper.

Next we show that the discretization proposed above preserves the conservation laws exactly. We start by noting that the total energy, total momentum, and density can be written as
\begin{align}\label{eq:energy-momentum-density}
E=\sum_{i\in I}e^iM(\psi_i, f_h), && \bm{P}=\sum_{i\in I}\bm{v}^i M(\psi_i, f_h), && N=\sum_{i\in I}1^i M(\psi_i, f_h),
\end{align}
where the coefficients $\{e^i\}_{i\in I}$, $\{\bm{v}^i\}_{i\in I}$, and $\{1^i\}_{i\in I}$ correspond to the degrees of freedom, or more precisely, to the expansion coefficients with respect to the chosen Galerkin basis for the functions $(|\bm{v}|^2,\bm{v},1)$, i.e.,
\begin{align}
|\bm{v}|^2=\sum_{i\in I} e^i\psi_i(\bm{v}) &&\bm{v}&=\sum_{i\in I} \bm{v}^i\psi_i(\bm{v}), && 1=\sum_{i\in I} 1^i\psi_i(\bm{v}).
\end{align}
This follows from the request that the basis $\{\psi_i\}_{i\in I}$ exactly reproduces quadratic functions. Also note that these coefficients are unique for any polynomial Galerkin basis. Hence the time derivatives of energy, momentum, and density vanish identically
\begin{align}
&\frac{dE}{dt}=\sum_{i,j\in I}e^i M(\psi_i, f_h \psi_j) \frac{dg^j}{dt}=\sum_{j\in I} C_{f_h}\left(\sum_{i\in I} e^i\psi_i , \psi_j\right)g^j=0,\\
&\frac{d\bm{P}}{dt}=\sum_{i,j\in I}\bm{v}^i M(\psi_i, f_h \psi_j) \frac{dg^j}{dt}=\sum_{j\in I} C_{f_h}\left(\sum_{i\in I} \bm{v}^i\psi_i , \psi_j\right)g^j=0,\\
&\frac{dN}{dt}=\sum_{i,j\in I}1^i M(\psi_i, f_h \psi_j) \frac{dg^j}{dt}=\sum_{j\in I} C_{f_h}\left(\sum_{i\in I} 1^i\psi_i , \psi_j\right)g^j=0,
\end{align}
where we have used the equations of motion for the degrees of freedom~\eqref{eq:Galerkin-matrix}, the bilinearity and the null-space of the form $C_f(u,w)$, and the requested property that the basis $\{\psi_i\}_{i\in I}$ reproduces quadratic functions exactly. 

To prove the existence and uniqueness of an equilibrium state, as well as the H-theorem, we first note that $\{e^i\}_{i\in I}$, $\{\bm{v}^i\}_{i\in I}$, and $\{1^i\}_{i\in I}$ are the only  eigenvectors of the matrix $C_{f_h}(\psi_i , \psi_j)$, that correspond to zero eigenvalues. This follows from the fact that, for the chosen second-order polynomial Galerkin basis, there exists only one unique set of coefficients, namely $\{e^i\}_{i\in I}$, $\{\bm{v}^i\}_{i\in I}$, and $\{1^i\}_{i\in I}$, in terms of which the null space $\phi=\{|\bm{v}|^2,\bm{v},1\}$ of the operator $C_f(u,w)$ can be expressed. 
%since the form $C_f(u,w)$, with $(u,w)\in \mathcal{W}$ and $\mathcal{W}$ an appropriate space such that $|C_f(u,w)| < \infty$, has a null space of $\phi=\{|\bm{v}|^2,\bm{v},1\}$ and $\{\psi_i\}_{i\in I}\subset \mathcal{W}$ is requested to contain $\phi$ with expansion coefficients corresponding to the vectors $\{e^i\}_{i\in I}$, $\{\bm{v}^i\}_{i\in I}$, and $\{1^i\}_{i\in I}$, there cannot be any other null eigenvectors for the matrix $C_{f_h}(\psi_i , \psi_j)$.
%\josh{This can't be right because $C_{ij}$ is a symmetric matrix, and therefore has a complete set of orthogonal eigenvectors. I think what we want is for the matrix to have no other \emph{null} eigenvectors. How do we prove that we have found all of the null eigenvectors? I understand how at the continuous level, but its not immediately obvious to me that the discretization preserves this property.}
%\eero{Fixed, I think.}
Hence the equilibrium state $g^i_{\text{eq}}$ must be a linear combination 
\begin{align}
g^i_{\text{eq}}=a\, e^i + \bm{b}\cdot\bm{v}^i+c\, 1^i,
\end{align}
which, within the support of our finite-element basis, corresponds to the numerical distribution function
\begin{align}
f_{h,\text{eq}}=\exp\left(a|\bm{v}|^2+\bm{b}\cdot\bm{v}+c\right).%, \quad \forall\, \bm{v}\in\text{supp}\left(\{\psi_i\}_{i\in I}\right).
\end{align}
Note that this expression is to be evaluated only within the supporting domain for the basis $\{\psi_i\}_{i\in I}$. Outside, it has no meaning.
%\josh{May be a good idea to mention that we are cutting off the velocity space somewhere. When I first read this I was bothered by the fact that outside the support of the finite element basis, the ansatz $f= \exp(S)$ implies that $f=1$. (This is because all of the $\psi_i$ are zero by definition outside the support of the basis, and $\exp(0)=1$.)}
%\eero{A remark added above. This is indeed a good point to stress.}
Furthermore, because the energy, momentum, and density are conserved, the coefficients $a$, $\bm{b}$, and $c$ are uniquely determined in terms of the moments of a given initial state. To conclude the H-theorem, we note that the entropy $S=-\int f_h\ln f_h d\bm{v}$ can be written as
\begin{align}\label{eq:finite-dimensional-entropy}
S=-\sum_{i\in I}M(f_h,\psi_i)g^i.
\end{align}
It's time derivative then becomes
\begin{align}
\frac{dS}{dt}&=-\sum_{i\in I}M(f_h,\psi_i)\frac{dg^i}{dt}-\sum_{i,j\in I}\frac{dg^j}{dt}M(\psi_jf_h,\psi_i)g^i\nonumber\\
&=-\frac{dN}{dt}-\sum_{i,j\in I}C_{f_h}(\psi_j,\psi_i)g^ig^j\nonumber\\
&\geq 0.
\end{align}
The last line follows from the density conservation and the fact that the form $C_f(u,w)$ is negative semidefinite, with the only nontrivial zero solution being a linear combination of the operator's null-space, corresponding to the equilibrium state. This concludes our proof for the direct Galerkin discretization.

Finally, we make some additional remarks. First, any generic map 
\begin{align}
f_h=\Psi(g_h), && g_h=\sum_ig^i(t)\psi_i(\bm{v}),
\end{align}
with $\Psi:\mathbb{R}\mapsto\mathbb{R}_{\geq 0}$ and combined with a polynomial basis of second order, would succeed in reproducing the conservation laws and the non-negativity constraint. %For example, in~\cite{Hirvijoki:2017ei} the map $x\mapsto\Phi(x) = x$, which is not positivity preserving,
%\josh{This is not a map $\mathbb{R}\rightarrow\mathbb{R}_{\geq  0}$. I thought [11] did not preserve positivity?}
%\eero{Fixed, I removed the reference as confusing}
%was used to obtain exact conservation laws. 
Second, any Galerkin basis that uniquely contains the functions $\phi(\bm{v})=\{|\bm{v}|^2,\bm{v},1\}$ within the support of the basis would suffice to provide the desired properties. For numerical purposes, it is convenient to use bases with compact support, though, as~\eqref{eq:Galerkin-matrix} is implicit. The exception would be a basis for which $M(\psi_i,f_h\psi_i)$ would become diagonal. Lastly, if the dyad $\mathbb{Q}$ were replaced with something more complicated, as in the case of a relativistic collision operator, all of the above could be generalized as long as the chosen Galerkin basis would exactly and uniquely reproduce the null-space of the corresponding $C_f(u,w)$, within the domain of support for the basis.

\section{Metriplectic structure of the direct Galerkin discretization}
\label{sec:metriplectic-structure}
At first, it seems as if the properties of our discretization scheme were a pure coincidence. There exists, however, a deeper level to the collision operator and its structure preserving discretizations. Specifically, the nonlinear Landau collision operator can be cast into an infinite-dimensional metriplectic system where the conservation laws correspond to so-called Casimir invariants~\cite{Morrison:1986vw}. Consequently, a careful discretization of the structure provides a finite-dimensional metriplectic system. As it happens, the discretization described previously ``accidentally'' defines a finite-dimensional metriplectic structure, and that the very same structure can be derived systematically, by discretizing the infinite-dimensional metriplectic structure. In this section, we demonstrate the correspondence of the direct Galerkin discretization to a finite-dimensional metriplectic system.

Let us start by multiplying equation~\eqref{eq:Galerkin-matrix} with the inverse of the matrix $M(\psi_i, f_h \psi_j)$, which leads to 
\begin{align}
\frac{dg^k}{dt} = \sum_{i,j\in I}M^{-1}(\psi_k, f_h \psi_i)\,C_{f_h}(\psi_i , \psi_j)\,g^j, \quad \forall\, k\in I.
\end{align}
Next we use the finite-dimensional entropy~\eqref{eq:finite-dimensional-entropy}, compute it's derivative with respect to $g^{\ell}$, and invert for the vector %providing us with
%\begin{align}
%\frac{\partial S}{\partial g^{\ell}}=-\sum_{j\in I} M(\psi_{\ell},f_h\psi_j)\,(g^j+1^j), \quad \forall\, \ell\in I.
%\end{align}
%Then we multiply the gradient of the entropy with the inverse of $M(\psi_i, f_h \psi_j)$, and solve for the vector 
\begin{align}\label{eq:entropy-derivative}
g^j+1^j=-\sum_{\ell\in I} M^{-1}(\psi_{j},f_h\psi_{\ell})\frac{\partial S}{\partial g^{\ell}}, \quad \forall\, j\in I.
\end{align}
In the next step, we use the fact that the vector $1^j$ is an eigenvector of the matrix $C_{f_h}(\psi_i,\psi_j)$ with a zero eigenvalue. This provides us the equations of motion in the form 
\begin{align}\label{eq:metriplectic-equation-of-motion}
\frac{dg^k}{dt} = -\sum_{\ell\in I}G_{k\ell}(g)\frac{\partial S}{\partial g^{\ell}}, \quad \forall\, k\in I,
\end{align}
where we have collected the individual matrices together and defined
\begin{align}\label{eq:metric-tensor}
G_{k\ell}(g)=\sum_{i,j\in I} M^{-1}(\psi_k, f_h \psi_i)\,C_{f_h}(\psi_i , \psi_j)\, M^{-1}(\psi_{j},f_h\psi_{\ell}).
\end{align}
To reveal the metriplectic structure in it's full glory, we consider time derivatives of generic functions $U(g)$ that depend only on the degrees-of-freedom $g=\{g^k\}_{i\in I}$, so that 
\begin{align}
\frac{d U(g)}{dt}=\sum_{k\in I}\frac{\partial U}{\partial g^k}\frac{d g^k}{dt}.
\end{align}
With the help of this identity, we may cast~\eqref{eq:metriplectic-equation-of-motion} into 
\begin{align}\label{eq:metriplectic-structure}
\frac{d U}{dt} = -\sum_{k,\ell\in I}\frac{\partial U}{\partial g^k} G_{k\ell}(g)\frac{\partial S}{\partial g^{\ell}}, \quad \forall\, k\in I,
\end{align}

The metriplectic structure of~\eqref{eq:metriplectic-structure} then follows from the facts that (i) the matrix $G_{k\ell}(g)$ is symmetric and negative semidefinite, (ii) it has the finite-dimensional energy, momentum, and density (defined in \eqref{eq:energy-momentum-density}) as Casimir invariants due to the conditions 
\begin{align}
\sum_{i\in I}\frac{\partial E(g)}{\partial g^i}\,G_{ij}(g)=0, &&
\sum_{i\in I}\frac{\partial \bm{P}(g)}{\partial g^i}\,G_{ij}(g)=0, &&
\sum_{i\in I}\frac{\partial N(g)}{\partial g^i}\,G_{ij}(g)=0, \quad \forall\, j\in I,
\end{align}
and that (iii) it has a unique equilibrium state
\begin{align}
g^i_{\text{eq}}=a\,e^i+\bm{b}\cdot\bm{v}^i+c\,1^i,
\end{align}
with the coefficients $a$, $\bm{b}$, and $c$ defined from the initial state, and that the equilibrium state satisfies the condition
\begin{align}
\sum_{i\in I}\frac{\partial S(g_{\text{eq}})}{\partial g^i}\,G_{ij}(g_{\text{eq}})=0, \quad \forall\, j\in I,
\end{align}
To complete the cycle, we next demonstrate how~\eqref{eq:metriplectic-structure}, and consequently~\eqref{eq:Galerkin-matrix}, can be obtained directly from discretizing the underlying infinite-dimensional metriplectic structure of the Landau collision operator.

\section{Discretization of the infinite-dimensional metriplectic structure}
\label{sec:metriplectic-discretization}
The infinite-dimensional metriplectic structure of the nonlinear Landau collision operator has been known for quite some time~\cite{Morrison:1986vw}. Here we explain its connection to the weak formulation~\eqref{eq:weakform} and demonstrate how discretization of it provides the same equations of motion as the direct Galerkin discretization introduced in Sec.~\ref{sec:direct-Galerkin}, when the same discretization is assumed for the distribution function as there.

In terms of a negative semidefinite, symmetric bracket defined according to
\begin{align}
(\fa,\fb)[f]=C_f\left(\frac{\delta\fa}{\delta f},\frac{\delta\fb}{\delta f}\right),
\end{align}
and an entropy functional defined as
\begin{align}
\fs[f]=M(-\ln f,f),
\end{align}
the dynamics of $f$ that reproduce~\eqref{eq:landau} can be recovered from the functional differential equation
\begin{align}\label{eq:infinite-dimensional-motion}
\frac{d\fu[f]}{dt}=(\fu,-\fs),
\end{align}
by choosing a functional $\fu[f]=M(u,f)$, and requiring the resulting equation to hold for all time-independent $u(\bm{v})$. This is left as an exercise for the reader to verify. Furthermore, entropy $\fs[f]$, energy $\fe[f]=M(|\bm{v}|^2,f)$, momentum $\bm{\fp}[f]=M(\bm{v},f)$, and density $\fn[f]=M(1,f)$ trivially satisfy
\begin{align}
\frac{dS}{dt}\geq 0, && (\fe,\fa)=0, && (\fp,\fa)=0, && (\fn,\fa)=0,
\end{align}
for arbitrary functionals $\fa$, and the equilibrium state corresponding to $d\fs/dt=0$ is achieved if and only if the functional derivative of the entropy is a linear combination of the functional derivatives of the Casimirs $\fe$, $\fp$, and $\fn$, corresponding to a Maxwellian. 

Before we proceed with the discretization, it is useful to investigate what happens if we use a map $f=\Psi(g)$. Since
\begin{align}\label{eq:modified-functional-derivative}
\frac{\delta \fa[f]}{\delta f}=\frac{1}{\Psi'(g)}\frac{\delta \fa[\Psi(g)]}{\delta g},
\end{align}
A metric bracket with respect to the function $g$ is obtained after the substitution
\begin{align}
(\fa,\fb)[g]=C_{\Psi(g)}\left(\frac{1}{\Psi'(g)}\frac{\delta \fa}{\delta g},\frac{1}{\Psi'(g)}\frac{\delta \fb}{\delta g}\right).
\end{align}
Our goal is to restrict $g$ to live within some finite-dimensional function space, basically choosing
\begin{align}
g_h(\bm{v},t)=\sum_{i\in I} g^i(t)\psi_i(\bm{v}),
\end{align}
and still preserve the Casimirs and the correct equilibrium state. This indicates that the expression~\eqref{eq:modified-functional-derivative} should be chosen so that it becomes exact for the Casimirs and the entropy when evaluated with respect to $g_h$, so that the null-space of $C_f(u,w)$ can be exploited. For the Casimirs, this issue has been discussed in detail in~\cite{Kraus-Hirvijoki-2017,Hirvijoki-Kraus-Burby:2018arXiv}. To obtain the exactness also for the entropy functional corresponding to Maxwell-Boltzmann statistics, it turns out that the choice $\Psi(x)=\exp(x)$ is essential. With these guidelines, following~\cite{Kraus-Hirvijoki-2017,Hirvijoki-Kraus-Burby:2018arXiv} then leads to the conclusion that
\begin{align}
\frac{\delta \fa[f_h]}{\delta f}=\frac{1}{\exp(g_h)}\frac{\delta \fa[\exp(g_h)]}{\delta g}=\sum_{i,j\in I}\frac{\partial \fa_h(g)}{\partial g^i}M^{-1}(\psi_i,f_h\psi_j)\psi_j,
\end{align}
where $\fa_h(g)=\fa[f_h]=\fa[\exp(g_h)]$. The finite-dimensional bracket with respect to the degrees of freedom $\{g^i\}_{i\in I}$ is then obtained by substituting the discrete functional derivative to the infinite-dimensional bracket providing 
\begin{align}
(\fa_h,\fb_h)_h(g)=(\fa,\fb)[f_h]=\sum_{k,\ell\in I}\frac{\partial \fa_h(g)}{\partial g^k} G_{k\ell}(g)\frac{\partial \fb_h(g)}{\partial g^{\ell}},
\end{align}
where the matrix $G_{k\ell}(g)$ is the one defined in~\eqref{eq:metric-tensor}. Finally, using~\eqref{eq:infinite-dimensional-motion}, we obtain
\begin{align}
\frac{d\fu_h(g)}{dt}=(\fu_h,-\fs_h)_h(g)=-\sum_{k,\ell\in I}\frac{\partial \fu_h}{\partial g^k} G_{k\ell}(g)\frac{\partial \fs_h}{\partial g^{\ell}},
\end{align}
which is the same result as given in~\eqref{eq:metriplectic-structure} (simply replace $\fu_h(g)$ with $U(g)$ etc.). All the properties are hence proven the same way as before. This concludes our discretization of the infinite-dimensional metriplectic structure.

%\josh{Wait, don't you want to write the metriplectic formulation after the change of variable $f=\exp(S)$?}
%\eero{Fixed.}

\section{Temporal discretization}
\label{sec:temporal}
Thus far, we have managed to convert the infinite-dimensional Landau collision operator to a finite-dimensional, time-continuous ordinary differential equation which has been shown to respect all of the properties present in the infinite-dimensional system. As a next step, we propose a novel integration method, that translates all of these desired properties to discrete time, providing a fully conservative and thermodynamically consistent set of equations that can be implemented on a computer. While recently there has been an ambitious attempt towards integration methods for metriplectic systems analoguous of symplectic integrators for Hamiltonian systems~\cite{Ottinger:2018}, we take a different route here.

To begin, we introduce the so-called discrete gradient methods~\cite{Quispel:1996,McLachlan:1999,CohenHairer:2011}, that are often times used to construct integrators for Hamiltonian systems while numerically preserving first integrals, e.g., the Hamiltonian, to machine precision. Given an ordinary differential equation of the form
\begin{align}
\frac{dg^k}{dt} = -\sum_{\ell\in I}G_{k\ell}(g)\frac{\partial S}{\partial g^{\ell}}, \quad \forall\, k\in I,
\end{align}
and denoting time instances with subscripts according to $g(\delta t)=g_1$ and $g(0)=g_0$, discrete gradient methods temporally approximate the above ODE system according to
\begin{align}\label{eq:temporal-discretization}
\frac{g^k_1-g^k_0}{\delta t}=-\sum_{\ell\in I}\overline{G}_{k\ell}[g_0,g_1]\overline{\frac{\partial S}{\partial g^{\ell}}}[g_0,g_1], \quad \forall\, k\in I.
\end{align}
The operator $\overline{\partial A/\partial g^{\ell}}[g_0,g_1]$ is referred to as the discrete gradient, and it is required to satisfy the properties
\begin{align}\label{eq:discrete-gradient-definition}
\sum_{\ell\in I}\overline{\frac{\partial A}{\partial g^{\ell}}}[g_0,g_1]\, (g_1^{\ell}-g_0^{\ell})=A(g_1)-A(g_0), &&
\overline{\frac{\partial A}{\partial g^{\ell}}}[g,g]=\frac{\partial A}{\partial g^{\ell}}(g).
\end{align}
Many such operators are known in the literature~\cite{HartenLaxLeer:1983,Gonzalez:1996}.
Furthermore, requiring $\overline{G}_{k\ell}(g,g)=G_{k\ell}(g)$ guarantees that the limit $\delta t\rightarrow 0$ collapses Eq.~\eqref{eq:temporal-discretization} to the correct time-continuous ordinary differential equation. 

Proceeding, and using the definition of the discrete gradient~\eqref{eq:discrete-gradient-definition} as well as the time-discrete evolution equation~\eqref{eq:temporal-discretization}, we note that the temporally-discrete evolution of any function $U(g)$ now satisfies 
\begin{align}
U(g_1)-U(g_0)=-\delta t \sum_{k,\ell\in I}\overline{\frac{\partial U}{\partial g^{k}}}[g_0,g_1]\overline{G}_{k\ell}[g_0,g_1]\overline{\frac{\partial S}{\partial g^{\ell}}}[g_0,g_1].
\end{align}
Hence, as long as the matrix operator $\overline{G}_{k\ell}[g_0,g_1]$ is negative semidefinite, entropy production will be guaranteed, according to
\begin{align}
S(g_1)-S(g_0)=-\delta t\sum_{k,\ell\in I}\overline{\frac{\partial S}{\partial g^{k}}}[g_0,g_1]\overline{G}_{k\ell}[g_0,g_1]\overline{\frac{\partial S}{\partial g^{\ell}}}[g_0,g_1]\geq 0.
\end{align}

Next we note the important result that will hint us on how to define the operator $\overline{G}_{k\ell}[g_0,g_1]$. For all Casimirs $C=\{E,\bm{P},N\}$, i.e., energy, momentum, and density, the derivative with respect to the degrees of freedom can be written in a convenient form, namely
\begin{align}
\frac{\partial C}{\partial g^k}=\sum_{k\in I}c^i M(\psi_i,f_h\psi_k),
\end{align}
with $c^i=\{e^i,\bm{v}^i,1^i\}$. A discrete gradient of the Casimirs is thus defined according to
\begin{align}
\overline{\frac{\partial C}{\partial g^i}}[g_0,g_1]=\sum_{k\in I}c^k \overline{M}_{ki}[g_0,g_1],
\end{align}
where $\overline{M}_{ki}[g_0,g_1]$ is required to satisfy the condition $\overline{M}_{ki}[g,g]=M(\psi_i,f_h\psi_k)$. The specific form of the matrix $\overline{M}_{ij}[g_0,g_1]$ depends on the chosen discrete gradient. We will provide a particularly convenient, explicit form soon. With these remarks, we see that the temporally-discrete evolution of the Casimirs satisfies
\begin{align}
C(g_1)-C(g_0)=-\delta t\sum_{i\in I}\sum_{k,\ell\in I}c^{i} \overline{M}_{i k}[g_0,g_1]\overline{G}_{k\ell}[g_0,g_1]\overline{\frac{\partial S}{\partial g^{\ell}}}[g_0,g_1].
\end{align}
If we are to achieve the discrete-time Casimir invariance $C(g_1)-C(g_0)=0$ for all possible state vectors $(g_1,g_0)$, we must choose
\begin{align}\label{eq:approximate-G}
\overline{G}_{k\ell}[g_0,g_1]=\sum_{i,j\in I} \overline{M}^{-1}_{ki}[g_0,g_1]\,C_{f_{h,1/2}}(\psi_i , \psi_j)\, \overline{M}^{-1}_{j\ell}[g_0,g_1],
\end{align}
where $\overline{M}^{-1}_{ij}[g_0,g_1]$ is the inverse of the matrix $\overline{M}_{ij}[g_0,g_1]$. This is also a valid choice: since we required $\overline{M}_{ij}[g,g]=M(\psi_i,f_h\psi_j)$, we trivially have $\overline{M}^{-1}_{ij}[g,g]=M^{-1}(\psi_i,f_h\psi_j)$, and hence $\overline{G}_{k\ell}(g,g)=G_{k\ell}(g)$. The choice~\eqref{eq:approximate-G} then provides the desired result
\begin{align}
C(g_1)-C(g_0)=-\delta t\sum_{i\in I}\sum_{j,\ell\in I}c^{i} C_{f_{h,1/2}}(\psi_i , \psi_j)\, \overline{M}^{-1}_{j\ell}[g_0,g_1]\overline{\frac{\partial S}{\partial g^{\ell}}}[g_0,g_1]=0,
\end{align}
which follows from the property that the basis $\{\psi_i\}_{i\in I}$ can present the functions $\phi=\{|\bm{v}|^2,\bm{v},1\}$ exactly, and due to the null space of the form $C_f(u,w)$, which together lead to
\begin{align}
\sum_{i\in I}c^{i} C_{f_{h}}(\psi_i , \psi_j)=0,  \quad \forall\, j\in I.
\end{align}

The final step is to verify the existence of a unique equilibrium sate. For an equilibrium state to exist, one must have $g_1=g_0=g_{\text{eq}}$. This requirement, and the evolution equation~\eqref{eq:temporal-discretization}, provides
\begin{align}
\sum_{\ell\in I}\overline{G}_{k\ell}[g_{\text{eq}},g_{\text{eq}}]\overline{\frac{\partial S}{\partial g^{\ell}}}[g_{\text{eq}},g_{\text{eq}}]=0, \quad \forall\, k\in I.
\end{align}
Next, using the defining properties $\overline{\partial S/\partial g^{\ell}}[g_{\text{eq}},g_{\text{eq}}]=\partial S/\partial g^{\ell}(g_{\text{eq}})$ and 
$\overline{G}_{k\ell}[g_{\text{eq}},g_{\text{eq}}]=G_{k\ell}(g_{\text{eq}})$, we obtain
\begin{align}
\sum_{\ell\in I}G_{k\ell}(g_{\text{eq}})\frac{\partial S}{\partial g^{\ell}}(g_{\text{eq}})=0, \quad \forall\, k\in I.
\end{align}
From here the uniqueness of the equilibrium state follows trivially after using~\eqref{eq:entropy-derivative} and the null-space argument, leading to the observation that the numerical equilibrium state is given by
\begin{align}
f_{h,\text{eq}}(\bm{v})=\exp(a|\bm{v}|^2+\bm{b}\cdot\bm{v}+c),
\end{align} 
as expected.

\section{The sparse matrix system}
To conclude our derivations, we choose a convenient discrete gradient method, and provide explicit expressions for all necessary terms, simultaneously converting~\eqref{eq:temporal-discretization} into a sparse form suitable for iterative inversion techniques.

We will use the second order $\mathcal{O}(\delta t^2)$, so-called average discrete gradient~\cite{HartenLaxLeer:1983}, that is defined according to 
\begin{align}
\overline{\frac{\partial A}{\partial g^{\ell}}}[g_0,g_1]=\int\limits_0^1\frac{\partial A}{\partial g^{\ell}}((1-\xi)g_0+\xi g_1)d\xi.
\end{align}
The explicit expression for the matrix $\overline{M}_{ij}[g_0,g_1]$ then becomes
\begin{align}
\overline{M}_{ij}[g_0,g_1]=\int \psi_i\frac{\exp\left(g_{h0}\right)-\exp\left(g_{h1}\right)}{g_{h0}-g_{h1}}\psi_j\,d\bm{v},
\end{align}
and the average discrete gradient of the entropy is given by
\begin{align}
\overline{\frac{\partial S}{\partial g^{\ell}}}[g_0,g_1]=\overline{\frac{\partial S - 1}{\partial g^{\ell}}}[g_0,g_1]-\overline{M}_{\ell j}[g_0,g_1]\,1^j,
\end{align}
where the vector $\overline{\partial S - 1/\partial g^{\ell}}[g_0,g_1]$ is defined as
\begin{align}
\overline{\frac{\partial S - 1}{\partial g^{\ell}}}[g_0,g_1]=-\int \psi_{\ell}\frac{(g_{h0}-1)\exp\left(g_{h0}\right)-(g_{h1}-1)\exp\left(g_{h1}\right)}{g_{h0}-g_{h1}}\,d\bm{v}.
\end{align}
For practical reasons, we have introduced the short notation
\begin{align}
g_{h0}=\sum_{k\in I}g_0^k\psi_k, && g_{h1}=\sum_{k\in I}g_1^k\psi_k.
\end{align}
Putting everything together, and using the null-space condition once more to remove the coefficients~$1^j$ in the gradient of entropy, we obtain a coupled sparse-matrix system,
\begin{align}\label{eq:sparse-system}
\sum_{k\in I}\overline{M}_{ik}[g_0,g_1]\frac{g^k_1-g^k_0}{\delta t}&=-\sum_{j\in I}C_{f_{h,1/2}}(\psi_i , \psi_j)F_{j}, \quad \forall\, i\in I,\\
\sum_{j\in I}\overline{M}_{ij}[g_0,g_1]F_{j}&=\overline{\frac{\partial S - 1}{\partial g^{i}}}[g_0,g_1], \quad \forall\, i\in I,
\end{align}
where $f_{h,1/2}=\exp((g_{h0}+g_{h1})/2)$. This system provides a nonlinear but sparse equation for solving $g_1$ in terms of $g_0$ (the dense matrix--vector product $C_{f_{h,1/2}}(\psi_i , \psi_j)F_{j}$ can be factored into sparse matrix--vector products as in~\cite{Kraus-Hirvijoki-2017}) and, if solved to machine precision, will provide all the desired properties to machine precision as well.

\section{Conclusions}
\label{sec:discussion}
In this article we have explored spatio-temporal discretizations for the nonlinear Landau collision operator. Our results consisted of three milestones: (i) the infinite-dimensional Landau collision operator was succesfully converted to an ODE with desired physical properties, (ii) this ODE was shown to have a metriplectic structure, and (iii) a successful temporal discretization of the metriplectic structure was achieved retaining all of the properties of the original infinite-dimensional system. Perhaps the most important observation was (ii), as without it, it would have been quite difficult to guess a successful temporal discretization scheme. Hence the milestone (ii) stresses the value of discretization methods that are based on the infinite-dimensional metriplectic formulation and, by preserving it's structure, annihilate the guesswork. Overall, the presented results constitute an important step on the path towards ever more refined structure-preserving discretization methods for the full Vlasov-Maxwell-Landau system.

\bibliographystyle{apsrev4-1}
\bibliography{bibfile}   

\providecommand{\noopsort}[1]{#1}
\begin{thebibliography}{16}%
\makeatletter
\providecommand \@ifxundefined [1]{%
 \@ifx{#1\undefined}
}%
\providecommand \@ifnum [1]{%
 \ifnum #1\expandafter \@firstoftwo
 \else \expandafter \@secondoftwo
 \fi
}%
\providecommand \@ifx [1]{%
 \ifx #1\expandafter \@firstoftwo
 \else \expandafter \@secondoftwo
 \fi
}%
\providecommand \natexlab [1]{#1}%
\providecommand \enquote  [1]{``#1''}%
\providecommand \bibnamefont  [1]{#1}%
\providecommand \bibfnamefont [1]{#1}%
\providecommand \citenamefont [1]{#1}%
\providecommand \href@noop [0]{\@secondoftwo}%
\providecommand \href [0]{\begingroup \@sanitize@url \@href}%
\providecommand \@href[1]{\@@startlink{#1}\@@href}%
\providecommand \@@href[1]{\endgroup#1\@@endlink}%
\providecommand \@sanitize@url [0]{\catcode `\\12\catcode `\$12\catcode
  `\&12\catcode `\#12\catcode `\^12\catcode `\_12\catcode `\%12\relax}%
\providecommand \@@startlink[1]{}%
\providecommand \@@endlink[0]{}%
\providecommand \url  [0]{\begingroup\@sanitize@url \@url }%
\providecommand \@url [1]{\endgroup\@href {#1}{\urlprefix }}%
\providecommand \urlprefix  [0]{URL }%
\providecommand \Eprint [0]{\href }%
\providecommand \doibase [0]{http://dx.doi.org/}%
\providecommand \selectlanguage [0]{\@gobble}%
\providecommand \bibinfo  [0]{\@secondoftwo}%
\providecommand \bibfield  [0]{\@secondoftwo}%
\providecommand \translation [1]{[#1]}%
\providecommand \BibitemOpen [0]{}%
\providecommand \bibitemStop [0]{}%
\providecommand \bibitemNoStop [0]{.\EOS\space}%
\providecommand \EOS [0]{\spacefactor3000\relax}%
\providecommand \BibitemShut  [1]{\csname bibitem#1\endcsname}%
\let\auto@bib@innerbib\@empty
%</preamble>
\bibitem [{\citenamefont {Landau}(1936)}]{Landau:1936}%
  \BibitemOpen
  \bibfield  {author} {\bibinfo {author} {\bibfnamefont {L.~D.}\ \bibnamefont
  {Landau}},\ }\href@noop {} {\bibfield  {journal} {\bibinfo  {journal}
  {Physikalische Zeitschrift der Sowjetunion}\ }\textbf {\bibinfo {volume}
  {10}},\ \bibinfo {pages} {154} (\bibinfo {year} {1936})}\BibitemShut
  {NoStop}%
\bibitem [{\citenamefont {Morrison}(1986)}]{Morrison:1986vw}%
  \BibitemOpen
  \bibfield  {author} {\bibinfo {author} {\bibfnamefont {P.~J.}\ \bibnamefont
  {Morrison}},\ }\href {\doibase 10.1016/0167-2789(86)90209-5} {\bibfield
  {journal} {\bibinfo  {journal} {Physica D: Nonlinear Phenomena}\ }\textbf
  {\bibinfo {volume} {18}},\ \bibinfo {pages} {410} (\bibinfo {year}
  {1986})}\BibitemShut {NoStop}%
\bibitem [{\citenamefont {Grmela}(1984)}]{Grmela:1984dn}%
  \BibitemOpen
  \bibfield  {author} {\bibinfo {author} {\bibfnamefont {M.}~\bibnamefont
  {Grmela}},\ }\href {\doibase 10.1016/0375-9601(84)90297-4} {\bibfield
  {journal} {\bibinfo  {journal} {Physics Letters A}\ }\textbf {\bibinfo
  {volume} {102}},\ \bibinfo {pages} {355} (\bibinfo {year}
  {1984})}\BibitemShut {NoStop}%
\bibitem [{\citenamefont {Quispel}\ and\ \citenamefont
  {Turner}(1996)}]{Quispel:1996}%
  \BibitemOpen
  \bibfield  {author} {\bibinfo {author} {\bibfnamefont {G.~R.~W.}\
  \bibnamefont {Quispel}}\ and\ \bibinfo {author} {\bibfnamefont {G.~S.}\
  \bibnamefont {Turner}},\ }\href {\doibase 10.1088/0305-4470/29/13/006}
  {\bibfield  {journal} {\bibinfo  {journal} {Journal of Physics A:
  Mathematical and General}\ }\textbf {\bibinfo {volume} {29}},\ \bibinfo
  {pages} {L341 } (\bibinfo {year} {1996})}\BibitemShut {NoStop}%
\bibitem [{\citenamefont {McLachlan}\ \emph {et~al.}(1999)\citenamefont
  {McLachlan}, \citenamefont {Quispel},\ and\ \citenamefont
  {Robidoux}}]{McLachlan:1999}%
  \BibitemOpen
  \bibfield  {author} {\bibinfo {author} {\bibfnamefont {R.~I.}\ \bibnamefont
  {McLachlan}}, \bibinfo {author} {\bibfnamefont {G.~R.~W.}\ \bibnamefont
  {Quispel}}, \ and\ \bibinfo {author} {\bibfnamefont {N.}~\bibnamefont
  {Robidoux}},\ }\href {\doibase 10.1098/rsta.1999.0363} {\bibfield  {journal}
  {\bibinfo  {journal} {Philosophical Transactions of the Royal Society of
  London A: Mathematical, Physical and Engineering Sciences}\ }\textbf
  {\bibinfo {volume} {357}},\ \bibinfo {pages} {1021} (\bibinfo {year}
  {1999})}\BibitemShut {NoStop}%
\bibitem [{\citenamefont {Cohen}\ and\ \citenamefont
  {Hairer}(2011)}]{CohenHairer:2011}%
  \BibitemOpen
  \bibfield  {author} {\bibinfo {author} {\bibfnamefont {D.}~\bibnamefont
  {Cohen}}\ and\ \bibinfo {author} {\bibfnamefont {E.}~\bibnamefont {Hairer}},\
  }\href {\doibase 10.1007/s10543-011-0310-z} {\bibfield  {journal} {\bibinfo
  {journal} {BIT Numerical Mathematics}\ }\textbf {\bibinfo {volume} {51}},\
  \bibinfo {pages} {91} (\bibinfo {year} {2011})}\BibitemShut {NoStop}%
\bibitem [{\citenamefont {{Buet}}\ and\ \citenamefont {{Le
  Thanh}}(2006)}]{Buet_LeThanh:hal-00092543}%
  \BibitemOpen
  \bibfield  {author} {\bibinfo {author} {\bibfnamefont {C.}~\bibnamefont
  {{Buet}}}\ and\ \bibinfo {author} {\bibfnamefont {K.-C.}\ \bibnamefont {{Le
  Thanh}}},\ }\href {https://hal.archives-ouvertes.fr/hal-00092543} {\enquote
  {\bibinfo {title} {{About positive, energy conservative and equilibrium state
  preserving schemes for the isotropic Fokker-Planck-Landau equation}},}\ }
  (\bibinfo {year} {2006}),\ \bibinfo {note} {working paper or
  preprint}\BibitemShut {NoStop}%
\bibitem [{\citenamefont {{Yoon}}\ and\ \citenamefont
  {{Chang}}(2014)}]{Yoon:2014:POP}%
  \BibitemOpen
  \bibfield  {author} {\bibinfo {author} {\bibfnamefont {E.~S.}\ \bibnamefont
  {{Yoon}}}\ and\ \bibinfo {author} {\bibfnamefont {C.~S.}\ \bibnamefont
  {{Chang}}},\ }\href {\doibase 10.1063/1.4867359} {\bibfield  {journal}
  {\bibinfo  {journal} {Physics of Plasmas}\ }\textbf {\bibinfo {volume}
  {21}},\ \bibinfo {eid} {032503} (\bibinfo {year} {2014})}\BibitemShut
  {NoStop}%
\bibitem [{\citenamefont {{Taitano}}\ \emph {et~al.}(2015)\citenamefont
  {{Taitano}}, \citenamefont {{Chac{\'o}n}}, \citenamefont {{Simakov}},\ and\
  \citenamefont {{Molvig}}}]{Taitano:2015JCP}%
  \BibitemOpen
  \bibfield  {author} {\bibinfo {author} {\bibfnamefont {W.~T.}\ \bibnamefont
  {{Taitano}}}, \bibinfo {author} {\bibfnamefont {L.}~\bibnamefont
  {{Chac{\'o}n}}}, \bibinfo {author} {\bibfnamefont {A.~N.}\ \bibnamefont
  {{Simakov}}}, \ and\ \bibinfo {author} {\bibfnamefont {K.}~\bibnamefont
  {{Molvig}}},\ }\href {\doibase 10.1016/j.jcp.2015.05.025} {\bibfield
  {journal} {\bibinfo  {journal} {Journal of Computational Physics}\ }\textbf
  {\bibinfo {volume} {297}},\ \bibinfo {pages} {357} (\bibinfo {year}
  {2015})}\BibitemShut {NoStop}%
\bibitem [{\citenamefont {{Hager}}\ \emph {et~al.}(2016)\citenamefont
  {{Hager}}, \citenamefont {{Yoon}}, \citenamefont {{Ku}}, \citenamefont
  {{D'Azevedo}}, \citenamefont {{Worley}},\ and\ \citenamefont
  {{Chang}}}]{Hager:2016:JCP}%
  \BibitemOpen
  \bibfield  {author} {\bibinfo {author} {\bibfnamefont {R.}~\bibnamefont
  {{Hager}}}, \bibinfo {author} {\bibfnamefont {E.~S.}\ \bibnamefont {{Yoon}}},
  \bibinfo {author} {\bibfnamefont {S.}~\bibnamefont {{Ku}}}, \bibinfo {author}
  {\bibfnamefont {E.~F.}\ \bibnamefont {{D'Azevedo}}}, \bibinfo {author}
  {\bibfnamefont {P.~H.}\ \bibnamefont {{Worley}}}, \ and\ \bibinfo {author}
  {\bibfnamefont {C.~S.}\ \bibnamefont {{Chang}}},\ }\href {\doibase
  10.1016/j.jcp.2016.03.064} {\bibfield  {journal} {\bibinfo  {journal}
  {Journal of Computational Physics}\ }\textbf {\bibinfo {volume} {315}},\
  \bibinfo {pages} {644} (\bibinfo {year} {2016})}\BibitemShut {NoStop}%
\bibitem [{\citenamefont {{Hirvijoki}}\ and\ \citenamefont
  {{Adams}}(2017)}]{Hirvijoki:2017ei}%
  \BibitemOpen
  \bibfield  {author} {\bibinfo {author} {\bibfnamefont {E.}~\bibnamefont
  {{Hirvijoki}}}\ and\ \bibinfo {author} {\bibfnamefont {M.~F.}\ \bibnamefont
  {{Adams}}},\ }\href {\doibase 10.1063/1.4979122} {\bibfield  {journal}
  {\bibinfo  {journal} {Physics of Plasmas}\ }\textbf {\bibinfo {volume}
  {24}},\ \bibinfo {eid} {032121} (\bibinfo {year} {2017})},\ \Eprint
  {http://arxiv.org/abs/1611.07881} {arXiv:1611.07881} \BibitemShut {NoStop}%
\bibitem [{\citenamefont {{Kraus}}\ and\ \citenamefont
  {{Hirvijoki}}(2017)}]{Kraus-Hirvijoki-2017}%
  \BibitemOpen
  \bibfield  {author} {\bibinfo {author} {\bibfnamefont {M.}~\bibnamefont
  {{Kraus}}}\ and\ \bibinfo {author} {\bibfnamefont {E.}~\bibnamefont
  {{Hirvijoki}}},\ }\href {\doibase 10.1063/1.4998610} {\bibfield  {journal}
  {\bibinfo  {journal} {Physics of Plasmas}\ }\textbf {\bibinfo {volume}
  {24}},\ \bibinfo {eid} {102311} (\bibinfo {year} {2017})},\ \Eprint
  {http://arxiv.org/abs/1707.01801} {arXiv:1707.01801} \BibitemShut {NoStop}%
\bibitem [{\citenamefont {{Hirvijoki}}\ \emph {et~al.}(2018)\citenamefont
  {{Hirvijoki}}, \citenamefont {{Kraus}},\ and\ \citenamefont
  {{Burby}}}]{Hirvijoki-Kraus-Burby:2018arXiv}%
  \BibitemOpen
  \bibfield  {author} {\bibinfo {author} {\bibfnamefont {E.}~\bibnamefont
  {{Hirvijoki}}}, \bibinfo {author} {\bibfnamefont {M.}~\bibnamefont
  {{Kraus}}}, \ and\ \bibinfo {author} {\bibfnamefont {J.~W.}\ \bibnamefont
  {{Burby}}},\ }\href@noop {} {\bibfield  {journal} {\bibinfo  {journal} {ArXiv
  e-prints}\ } (\bibinfo {year} {2018})},\ \Eprint
  {http://arxiv.org/abs/1802.05263} {arXiv:1802.05263} \BibitemShut {NoStop}%
\bibitem [{\citenamefont {{\"Ottinger}}(2018)}]{Ottinger:2018}%
  \BibitemOpen
  \bibfield  {author} {\bibinfo {author} {\bibfnamefont {H.}~\bibnamefont
  {{\"Ottinger}}},\ }\href {\doibase 10.1515/jnet-2017-0034} {\bibfield
  {journal} {\bibinfo  {journal} {{Journal of Non-Equilibrium Thermodynamics}}\
  }\textbf {\bibinfo {volume} {{42}}},\ \bibinfo {pages} {89} (\bibinfo {year}
  {2018})}\BibitemShut {NoStop}%
\bibitem [{\citenamefont {Harten}\ \emph {et~al.}(1983)\citenamefont {Harten},
  \citenamefont {Lax},\ and\ \citenamefont {Leer}}]{HartenLaxLeer:1983}%
  \BibitemOpen
  \bibfield  {author} {\bibinfo {author} {\bibfnamefont {A.}~\bibnamefont
  {Harten}}, \bibinfo {author} {\bibfnamefont {P.~D.}\ \bibnamefont {Lax}}, \
  and\ \bibinfo {author} {\bibfnamefont {B.~v.}\ \bibnamefont {Leer}},\ }\href
  {\doibase 10.1137/1025002} {\bibfield  {journal} {\bibinfo  {journal} {SIAM
  Review}\ }\textbf {\bibinfo {volume} {25}},\ \bibinfo {pages} {35} (\bibinfo
  {year} {1983})}\BibitemShut {NoStop}%
\bibitem [{\citenamefont {Gonzalez}(1996)}]{Gonzalez:1996}%
  \BibitemOpen
  \bibfield  {author} {\bibinfo {author} {\bibfnamefont {O.}~\bibnamefont
  {Gonzalez}},\ }\href {\doibase 10.1007/978-1-4612-1246-1_10} {\bibfield
  {journal} {\bibinfo  {journal} {Journal of Nonlinear Science}\ }\textbf
  {\bibinfo {volume} {6}},\ \bibinfo {pages} {449} (\bibinfo {year}
  {1996})}\BibitemShut {NoStop}%
\end{thebibliography}%
\end{document}